\documentstyle[aps,epsf]{revtex}  
\begin{document}
\title{The infinite-range quantum random 
Heisenberg magnet.}
\author{
Liliana Arrachea$^{1,2}$ and Marcelo J. Rozenberg$^{1}$} 
\address{$^1$ Departamento de F\'{\i }sica, \\
FCEyN Universidad de Buenos Aires\\
Pabell\'{o}n I, Ciudad Universitaria, (1428) Buenos Aires, Argentina.\\
$^2$ Scuola Internazionale Superiore di Studi Avanzati (SISSA) and 
Istituto Nazionale per la Fisica della Materia (INFM)\\
Unit\`a di Ricerca Trieste-SISSA,
Via Beirut 4, I-34014 Trieste, Italy.  }
\date{Received \today }
\maketitle

\begin{abstract}
We study with exact diagonalization techniques the Heisenberg model for a
system of SU(2) spins with $S=1/2$ and random infinite-range
exchange interactions.   
We calculate the critical temperature $T_g$ for the spin-glass
to paramagnetic transition. We obtain $T_g \approx 0.13$,
in good agreement with previous quantum Monte Carlo and analytical
estimates. 
We 
provide a detailed picture for
the different kind of excitations which intervene
in the dynamical response $\chi^{\prime \prime}(\omega,T)$
at $T=0$ and analyze  their evolution
as $T$ increases. We also calculate the specific heat $C_v(T)$.
We find that it displays a smooth maximum at $T_M \approx 0.25$, in good
qualitative agreement with experiments. We argue that the fact that
$T_M>T_g$ is due to a quantum disorder effect.  
\end{abstract}


\newpage

\section{Introduction}

The physics of disordered magnets is a fascinating subject of
condensed matter physics. 
Traditionally, two main ingredients are singled out as crucial to set 
the physical 
behavior of 
these systems: strong interaction and frustration. It is well known
that the interplay between them leads to
a rich variety of magnetically ordered phases, including 
conventional commensurate or incommensurate spin-density waves as
well as the more exotic spin-glass state. The latter is characterized
by an ordered magnetic state with permanent magnetic moments in the microscopic
scale, but randomly oriented producing a vanishing net macroscopic 
magnetization.
While the concept of spin itself is purely quantum, it is often maintained
that quantum fluctuations are not 
important for spin-glass physics \cite{fisher}. However, recent
experiments have put the role of quantum fluctuations \cite{subir} at the center of
the stage.

One example of real systems with a spin-glass phase at low temperature
is the compound Li$_{1-x}$Ho$_x$YF$_4$ which is a dipolar coupled random magnet
\cite{holmio} and has been recently the focus of  beautiful experiments
\cite{aeppli} that introduced quantum fluctuations by means of an
external transverse
magnetic field.   Another notable example is the LiV$_2$O$_4$ \cite{liv}
compound in  which the magnetic V atoms
are placed at the vertices of a phyrochlore-like structure which produces
strong frustration and a spin-glass state. 
The V $d$-electrons also form a very narrow conduction
band that behaves as a strongly renormalized Fermi liquid, with 
parameters comparable
to those of the so called heavy fermion compounds that usually involve only
$f-$electrons. The likely connection between this observation
and the spin-glass state remains a challenging open question.
Finally, we have the cuprate superconductors, with vast experimental 
evidence that a glassy
phase exists at low temperatures within a narrow range of doping 
concentrations 
between the antiferromagnetic and superconducting phases \cite{pdsup}.

Much effort has been dedicated to investigate the spin-glass physics,
and many ingenious and insightful theoretical ideas have contributed 
to our current 
understanding \cite{fisher,subir,young,binder,parisi}.
Nevertheless, despite this effort, many fundamental questions remain unsolved
and the role of quantum fluctuations is only beginning to receive due 
attention. 
Among the many long-standing unresolved issues of spin-glass physics 
we can mention
the intriguing behavior of the specific heat. In fact, experiments show that 
this quantity systematically has a maximum well above the spin-glass freezing
temperature $T_g$ \cite{fisher,binder}. 
It is usually claimed that quantum mechanical effects are
not essential for the physical phenomena related with the spin-glass
phase \cite{fisher}. However, we shall later argue that precisely quantum
effects might be at the origin of this long-standing puzzle.  

Theoretical progress in the description of the glassy phase is usually
prevented by technical difficulties.
In particular, most of the analytical and
numerical approaches
rely on the so called replica trick \cite{parisi}.
Unfortunately, this clever technique becomes 
usually impractical within the glassy phase when replica symmetry 
breaking occurs. In a recent paper \cite{limarprl} we introduced the use of
the method of numerical exact diagonalization of finite size clusters to 
investigate models of quantum random magnets at $T=0$.
This approach has two main advantages: (i) 
Averages over disorder can be directly performed, avoiding
the use of replicas. 
In fact, the same effort is required
to tackle both the disordered  and the spin-glass ordered phases. 
(ii) The dynamical response
is directly calculated on the real frequency axis and, unlike other
numerical techniques such as quantum Monte Carlo, no analytical 
continuation from the imaginary axis is necessary.  
Another advantage is that unlike usual exact diagonalization
calculations that give just a few poles in the response functions,
in our case we obtain smooth functions due to the average over the
disorder. 
However, the main drawback of this numerical 
method is that systems with a rather small number of spins are tractable. 
Nevertheless, this technical obstacle can be overcome by a careful
finite size analysis. In fact, we found that most of the relevant physical 
quantities exhibit a smooth behavior as a function of the system size,
and reliable extrapolations to the thermodynamic limit can be obtained
\cite{limarprl,conf}. Moreover, as we have complete knowledge of the system
for every realization, including the ground state wave function, we can
look at its structure to try to gain new insights. This has indeed
turned out to be a useful idea, as we obtain an appealing physical picture
of the low energy excitations of the spin-glass state, that would have
not emerged from
classical model numerical calculations.

In this work we consider the SU(2) Heisenberg model for a system
of $S=1/2$ spins with random infinite-range
exchange random interactions. It is defined by the hamiltonian
\begin{equation}
H = \frac{1}{\sqrt{N}} 
\sum_{i,j=1}^{N} J_{ij} {\bf S}_i {\bf \cdot S}_j,
\label{1}
\end{equation}
where $i,j$ labels sites of a lattice with $N$ spins, ${\bf S}$ denote the 
SU(2) 
spin $1/2$ operators and  the infinite-range exchange constants  
$J_{ij}$ are normally distributed with variance $J^2$ that we set to unity. 
 The phase diagram of this model was first outlined by Bray and Moore 
\cite{bm}, 
who argued that a glassy phase should  exist at low temperature for all values
of $S$. These authors proposed an approach which is formulated on the 
imaginary 
time axis and uses
the replica trick to obtain a set of self-consistent dynamical mean-field 
equations. The exact numerical solution of these
equations was later obtained with quantum Monte Carlo techniques 
in the paramagnetic phase \cite{daniel},
and the solution was found to become 
unstable towards spin-glass order
at a critical temperature $T_g \approx 0.14J$. 
The study of SU(M) extensions of this model
has been treated  in the limit of M$
\rightarrow \infty $ \cite{subirsum,georges,parcollet}, and spin-glass and
spin-liquid phases where found at sufficiently low temperatures.

We have recently investigated the dynamical response  
of the hamiltonian (\ref{1}) at zero temperature
with exact diagonalization techniques \cite{limarprl}.  
We were able to describe in detail the structure of the ground state
and the nature of the elemental excitations within the glassy phase.
 We found
compelling evidence that the dynamical spin response behaves
as $\chi_{loc}^{\prime \prime }(\omega )\sim {\rm q}\delta (\omega )+\chi
_{reg}(\omega )$ in the thermodynamic limit.
We
estimated $q \sim 0.06$ for the value of the Edwards-Anderson order parameter.

The aim of this paper is to extend the methodology based in exact 
diagonalization
to investigate the physical behavior of the disordered Heisenberg
model (\ref{1}) at finite temperature. Our main goal is to achieve a detailed
understanding of the different kind  of excitations which occur in the
dynamical spin response   at different 
temperatures, within and above the glassy phase. 
We shall show that the regular part 
$\chi_{reg}(\omega)$ is in fact dominated by  
spin excitations due to quantum disorder.
This contribution to the response function can be qualitatively
understood in the framework of an heuristic mean-field theory
that we present.
Moreover, its functional form is 
very similar to that of the quantum spin-liquid discussed in Refs.
\cite{subirsum,georges,parcollet} for the SU(M) generalization of the
Heisenberg model in the regime of large M and small S. 
We also investigate the
behavior of the specific heat $C_v$ as a function of temperature.
We find that this quantity displays a smooth maximum at a temperature
$T_M$ well above the freezing temperature $T_g$, and we link this fact
to the presence of strong quantum fluctuations. 
Interestingly, the unconventional behavior of the specific
heat is in accordance with the observation of this effect
in real materials that have a spin-glass state at low temperatures
and remained unexplained so far.

The paper is organized as follows. 
In Section II
we explain the numerical method and some technical details.
Section III contains the study of the dynamical spin response.
Results for the specific heat $C_v$ are shown and discussed in 
Section IV. The summary and conclusions are presented in the
Section V.

\section{Methodology.}
The general strategy is to
take samples from the random ensemble of systems of size $N$
and exactly diagonalize the ensuing hamiltonians (\ref{1}).
The different physical quantities are computed for each 
realization and then averaged over the number of samples.
Finite size effects are analyzed and results are extrapolated 
to the thermodynamic limit. Typically, systems with up to
$N=17$ spins are solved at $T=0$ and up to $N=12$ at finite $T$.
Averages are performed over
several thousands to hundreds of thousands of disorder 
realizations. A typical run
demands up to a week for the larger systems on an 8 node parallel
cluster.

The ground state and the dynamical correlation functions at $T=0$
are calculated by the Lanczos method \cite{lan}. 
It is convenient to take advantage of the SU(2) symmetry 
of the model.
The selected basis of states belongs to the $S^z=0$ representation, 
where $S^z$ is the quantum number corresponding to the
$z$-component of the total spin. 
A projector is used to find the ground state within each subspace 
with total spin $S$ \cite{lili}.  
The local spin susceptibility
is obtained from 
\begin{equation}
\chi_{loc}(\omega)=\frac{1}{M} \sum_{m=1}^{M} 
\frac{1}{N} \sum_{i=1}^{N}
\frac{1}{3} \sum_{\alpha} 
\langle \Phi^{(m)}_0 |S_i^{\alpha} \frac{1}{\omega - H^{(m)}}
S_i^{\alpha} |\Phi^{(m)}_0 \rangle,
\label{2}
\end{equation}
where $M$ is the number of realizations of disorder,  
$|\Phi^{(m)}_0\rangle$ denotes the ground state for the 
$J_{ij}$ set corresponding to the $m$-th realization, and
$\alpha=x,y,z$ labels the three components of 
the spin operator. Although we deal with systems having a finite number
of poles for each realization, the average over disorder naturally produces 
smooth response functions without need of introducing an artificial broadening
as in usual exact diagonalization methods. In some cases, we found useful
to use a logarithmic discretization of the $\omega$-axis to obtain 
accurate results due to the large number of poles occurring at low 
frequencies.   

To study the physical behavior at finite $T$, 
we
exactly diagonalized the full hamiltonian matrix
$H^{(m)}$. 
Each subspace belonging to the different $S^z$ representations is
separately diagonalized in order
to optimize the use of memory. 
In this case, the  local spin susceptibility
is obtained from the spectral function
\begin{eqnarray}
\chi^{\prime \prime}_{loc}(\omega) &=&\frac{1}{M}\sum_{m=1}^{M} 
\frac{1}{N} \sum_{i=1}^{N}
\frac{1}{Z^{(m)}}
\sum_{k,l} |\langle \Phi^{(m)}_k |S_i^z |\Phi^{(m)}_l \rangle|^2
[\exp(-\beta E^{(m)}_k) \delta(\omega-(E^{(m)}_k-E^{(m)}_l))
\nonumber \\
&-&
\exp(-\beta E^{(m)}_l) \delta(\omega-(E^{(m)}_l-E^{(m)}_k))],
\label{3}
\end{eqnarray}
where $\chi^{\prime \prime}_{loc}(\omega)=-2/\pi Im[\chi_{loc}(\omega)]$,
while 
$E^{(m)}_k$ is the eigenenergy of the eigenstate
$|\Phi^{(m)}_k \rangle$ corresponding to the 
$m$-th realization of disorder, $\beta=1/T$ is the inverse of the
temperature and $Z^{(m)}$ is the ensuing partition function.

The spin-glass phase is signaled by the divergence of the spin-glass
susceptibility $\chi_ {SG}$, which is related to the local-spin
susceptibility by  \cite{fisher} 
\begin{equation} 
\chi_{SG}=\frac{\chi_{loc}^2}{1-J^2 \chi_{loc}^2 },
\label{4}
\end{equation}  
where $\chi_{loc}=Re[\chi_{loc}(\omega=0)]$. Thus, the condition
\begin{equation}
J \chi_{loc}=1
\label{5}
\end{equation}
indicates the instability of the system towards a spin glass state.

In  previous papers \cite{limarprl,conf} it was demonstrated the accuracy of the method
by reproducing several known results
for the infinite-range Ising model with random exchange interactions and
transverse magnetic field $\Gamma$ \cite{mardan}. 
In particular, an accurate estimate for the critical value 
of the transverse field $\Gamma_c $,  at which a quantum transition
between the spin-glass and the paramagnetic phases takes place, 
was obtained.
At finite $T$  we should also test 
the accuracy of this approach for the current model we are studying.
There are not many well known results 
that can be used for benchmark of this numerical method. To the best of our
knowledge, the only quantitative prediction is
$T_g \approx 0.14 J $  for the critical temperature
at which the spin-glass to paramagnetic phase transition occurs. 
This estimate was obtained 
by quantum Monte Carlo numerical solution \cite{daniel}
of the mean-field equations derived by Bray and Moore \cite{bm}.

Results for the behavior of $\chi_{loc}$ as a function of $T$ are
shown in Fig. 1. The curve with circles correspond to 
extrapolations to the thermodynamic limit
obtained from data for systems with up to $N=12$ spins. 
At high temperatures, the local susceptibility obeys
a Curie law, as expected. The asymptotic behavior 
$\chi_{loc}=\beta /4 $ is plotted in dashed
lines for comparison. As the temperature  
decreases, quantum fluctuations cause the reduction of
the effective local magnetic moment and the response at 
$\omega=0$ becomes smaller than that of classical spins. 
The condition (\ref{5})
is satisfied at the critical temperature $T_g \approx 0.13$ where
the systems begin to freeze.
The diamond over the curve indicates the corresponding
result obtained with quantum Monte Carlo \cite{daniel}
that shows very good agreement and validates our approach
at finite $T$.

In Fig. 2 we show the behavior of  $\chi_{loc}$ as a function of $1/N$ for a selected 
set of temperatures. A linear fit was performed to obtain the extrapolated
values at $1/N=0$.
 Error bars in Fig. 1 indicate the corresponding
standard deviation. Interestingly, an even-odd effect is observed at 
temperatures below $T_g$, which suggests a change in the scaling law 
as the system enters the ordered phase.
    
\section{The dynamical spin response.}
\subsection{$\chi^{\prime \prime}_{loc}(\omega)$ at $T=0$.}
Let us begin with the study of the spin response at $T=0$.
In this subsection we shall summarize and extend our recent results
of Ref. \cite{limarprl}. 

The spectral function  $\chi_{loc}^{\prime \prime }(\omega )$
is plotted in Fig. 3 for systems of different sizes. 
As we discussed in our previous paper \cite{limarprl}, the most prominent
feature is  the existence of a $\sim
\delta (\omega )$ piece plus a low frequency hump
mounted on a regular contribution. Here, we shall present a more
detailed analysis of the various contributions to
$\chi_{loc}^{\prime \prime }(\omega )$, including some new excitations
that were not previously noticed. We shall present just
an heuristic, by no means rigorous, mathematical description of the 
different contributions to
$\chi_{loc}^{\prime \prime }$ that will provide us with an appealing physical
picture.

To gain insight on the nature 
of the dynamical response we analyzed
individual realizations of disorder. 
We distinguished four different contributions which we shall 
describe in detail throughout this section:
\begin{equation}
\chi_{loc}^{\prime \prime }(\omega )= K \delta (\omega ) +
\chi^{\prime \prime}_{low}(\omega )+\chi_{high}^{\prime \prime}(\omega )
+\chi^{\prime \prime}_{inc}(\omega ),
\label{6}
\end{equation}
where $\chi_{low}^{\prime \prime}(\omega )$ corresponds to
the low frequency hump, 
$\chi_{high}^{\prime \prime}(\omega )$ indicates a feature at high frequency 
${\cal O}(J)$ carrying a small spectral weight,
while $\chi_{inc}^{\prime \prime}(\omega )$ denotes the contribution of 
excitations
of incoherent nature which
conform a quantum disordered paramagnetic background. 

We will discuss the physical nature of all these contributions next.
We begin with the latter that is due to incoherent excitations.
We found that it can be described with
the following 
simple expression
\begin{equation}
\chi_{inc}^{\prime \prime}(\omega )=C \exp{[-\frac{\omega^2}{2 J^2 S(S+1)}]},
\label{7}
\end{equation}
with $S=1/2$ and $C$ a constant.
This function is plotted in thick line in Fig. 3 along with our results
for $\chi_{loc}^{\prime \prime }(\omega )$.

We now show that this contribution 
can be heuristically described as independent spins in the 
presence of effective randomly fluctuating magnetic fields ${\bf h}$.
We stress that we were not able to derive any proper demonstration
of the validity of this procedure and we include it here with the
sole hope of motivating the reader and perhaps providing new insights.
The fields are the molecular fields at each site due to the action
of all other spins,
\begin{equation}
{\bf h}=
\frac{1}{\sqrt {N}} \sum_{i,j} J_{ij} {\bf S}_j.
\label{mf}
\end{equation}
Following standard arguments \cite{fisher}, 
we assume that for a given realization of disorder, they point to any
spatial direction with equal normal probability
\begin{equation} 
P(h) \propto 
\exp[- \frac{h^2 }{2 \sigma^2}],
\label{prob}
\end{equation} 
 where
the variance $\sigma$ can be taken to be
$\sigma^2=\langle {\bf h \cdot h} \rangle \approx J^2 S(S+1)$.

We now recall that a spin under a magnetic field 
has a dynamical response given by 
\begin{equation}
\chi^{\prime \prime}_{o}(\omega,h)=\frac{1}{3}
 [\frac{1}{4}
\delta(\omega) + \frac{1}{2} \delta(\omega-h)],
\label{chimf}
\end{equation}
where the finite frequency transverse
part depends only on the magnitude of the field $h$.

Finally, we may combine (\ref{prob}) and (\ref{chimf}) to obtain
the incoherent background part of the dynamical
spin response function within this 
heuristic picture
\begin{equation}
\chi_{inc}^{\prime \prime}(\omega) 
\propto \int^{+\infty}_{-infty} dh  P(h) 
\chi_{o}^{\prime \prime}(\omega,h),
\label{tot}
\end{equation}
that leads to (\ref{7}) where 
the constant $C$ = 0.175 is set to fit the numerical data.
We carefully examined systems of
several sizes and found that $C$ varies negligible with $N$.
A point we would like to make is that the averaging (\ref{tot})
involves a one-dimensional integral. The reason is that we are describing
the quantum disordered regular contribution, therefore these fields cannot
be assumed to have a fixed direction in space.

Another interesting observation is that this contribution 
to the response function,
is very similar to one found
in the spin-liquid phase of the SU(M) generalization of the model
at large M and in the quantum disordered regime (small $S$)
\cite{subirsum,georges,parcollet}.

In order to study the remaining three pieces of 
$\chi_{loc}^{\prime \prime }(\omega )$ in more detail, we 
substract $\chi_{inc}^{\prime \prime}(\omega)$ from this quantity and define
\begin{equation}
\delta \chi^{\prime \prime }_{loc}(\omega ) = 
\chi_{loc}^{\prime \prime }(\omega )-
\chi^{\prime \prime }_{inc}(\omega).
\label{8}
\end{equation}
This quantity exhibits a strong dependence on the system size $N$ and
a careful finite size analysis is necessary in order to extract reliable
conclusions.

The $K \delta(\omega)$ part is a direct
consequence of the SU(2) rotational invariance of the Hamiltonian. 
To demonstrate this we sorted out the contributions to 
$\delta \chi_{loc}^{\prime \prime }(\omega )$ coming from the 
different $S$-sectors.
In Fig. 4 
we show results for the $S=0$, $S=1$ and $S=2$ sectors. A remarkable
observation is that the $\sim \delta (\omega )$ part is absent in the $S=0$
case, while present in the remaining $S\neq 0$ sectors. 
Moreover, the systematic analysis
of histograms of $S$ (cf. Fig. 5a) for increasing $N$, 
reveals that $\langle S \rangle \propto \sqrt{N}$ as shown in Fig. 5b.
Therefore, although the mean total spin $\langle S \rangle \neq 0$ in the
thermodynamic limit (in
fact diverges), the local magnetization, i.e. the magnetization {\em per site}
 $\langle S \rangle /N \rightarrow 0$. 
We have thus traced the origin of this $\sim \delta (\omega )$ feature to 
a ``soft mode'' connecting the $2S+1$-fold degenerate ground state 
that persists in the $N \rightarrow \infty $ limit.

The low frequency hump $\chi_{low}^{\prime \prime}(\omega)$ evolves 
 towards a narrowing feature at $\omega =0$ that becomes a $\delta$-function
in the thermodynamic limit. 
In fact, the spectral weight  of this part contributes together with
$K$ to the Edwards-Anderson order parameter $q$.  
To elucidate the origin of this hump, we looked carefully to the
structure of the ground state wave function for realizations of disorder that
yielded spectral weight in the energy range of the hump 
$\chi_{low}^{\prime \prime}(\omega)$.
We found that in those 
realizations
the ground state  wave function is dominated by some large
amplitudes of just a few pairs of states. The states conforming
a pair are related by a time-reversal
operation \cite{acla}. They appear in a symmetric (when $N/2$ is even) or
antisymmetric (when $N/2$ is odd) linear combination. Moreover, we also
observed that most of the
individual spin states in each of the two states of 
a given pair are in a configuration 
compatible with the
particular realization of $J_{ij}$ interactions. Thus we can think of this large 
fraction of spins as defining an unfrustrated cluster of size $N_c$
(${\cal O}(N)$). Therefore, several unfrustrated clusters are defined for
each realization of disorder. In fact, a number equal to the number of
dominating pair of amplitudes occurring in the ground state wave function.
An appealing picture to visualize the ground state is that the clusters
undergo a quantum mechanical tunneling back and forth between the two
corresponding pair of states. As many pairs coexist in any given
ground state, then the picture is that many unfrustrated cluster
simultaneously undergo these quantum oscillations.

With this physical picture of the ground state
in mind it should not be hard to anticipate
that the excitations that contribute to 
$\chi_{low}^{\prime \prime}(\omega)$ (the hump) are nothing but
wave functions where one (any) of the clusters has 
its pair of associated amplitudes appearing with the opposite
symmetry respect to that in the ground state (a ``flipped''
cluster). Since many coexistent clusters
occur in a single ground state for a given realization, then there are many
corresponding excited states contributing to the hump.
We can think of these as real collective excitations, as they involve
the simultaneous change in the state of a large number of individual
spins.
We have confirmed this qualitative picture by direct inspection of the amplitudes 
of the wave functions of the excited states in many realizations of disorder.

An important observation that we made is that
the difference in energy between
the ground state and the excitations is small (we found it to be
${\cal O}(1/N)$).  Thus, these collective excitations
become degenerate with the ground state and contribute with a sizable weight
to  the dynamical spin response at $\omega =0$ in the thermodynamic limit.
In other words, the hump collapses into a $\delta$-function contribution
in that limit. 
A side consequence of this observation is that even within the S=0 sector,
which also shows a hump in the dynamical response,
the T=0 state is that of a spin-glass with local magnetizations
with long range order in time.

To perform a quantitative analysis of the behavior
of $\chi_{low}^{\prime \prime}(\omega)$ as a function of $N$, we used 
the following parametrized form
\begin{equation}  
\chi_{low}^{\prime \prime}(\omega)= A \exp[-\frac{\omega^2}{\Gamma^ 2}].
\label{9}
\end{equation}
and tracked the systematic evolution of the parameters as function
of temperature and system size.

Typical curves are shown in 
Fig. 6 for two different sizes. The fitting parameters $A$ and $\Gamma$ 
have a strong dependence on the system size. Their behavior as functions
of $1/N$ is shown in Fig. 7. The width $\Gamma$ obeys a linear extrapolation
to $0$, while the integral of $\chi_{low}^{\prime \prime}(\omega)$ 
(squares in Fig. 7) 
remains approximately constant. This provides further quantitative
support to the claim
that $\chi_{low}^{\prime \prime}(\omega)$ evolves to a $ \sim \delta(\omega)$
contribution in the thermodynamic limit. 
The integrated spectral weight provides an estimate for the Edwards-Anderson
order parameter $q$. With this procedure it is found $q \sim 0.04$, which
improves upon our previous estimate $q \sim 0.06$ 
\cite{limarprl}. We shall later discuss the origin of this difference. 

To complete this subsection, we turn to discuss a last and small contribution
to $\chi_{loc}^{\prime \prime}(\omega)$, the high frequency part
$\chi_{high}^{\prime \prime}(\omega)$. This piece of the spectral function is 
related with the existence of a spin-glass order and it is essentially
the high energy counterpart of $\chi_{low}^{\prime \prime}(\omega)$.
Similar to the former case, $\chi_{high}^{\prime \prime}(\omega)$
is due to excitations of the clusters. But the key difference is that
now the excitations are incoherent in nature,
involving the unbinding of single spins out of the unfrustrated
clusters. The spins will then independently revolve around its local field
with a fast precession frequency. 
In other words, the relevant excitations contributing to 
$\chi_{high}^{\prime \prime}(\omega)$ correspond 
to the flipping of individual spins out of the large unfrustrated clusters. 

An heuristic mean field picture can also be presented
in this case.
To
describe these processes let us think of an
originally unfrustrated cluster of $N_c$ spins as being
made of a single spin plus the remaining cluster
of $N_c-1$ correlated spins.   
The latter would produce an effective magnetic field pointing to some fixed 
(or very slowly moving)
direction which couples to the single spin that
will undergo a fast precession
around the slow effective field due to the 
cluster.  
To estimate
the magnitude of the effective field 
$h=|{\bf h}|$, we assume it to be normally distributed around 
$h_0 = | \sum_{ij} J_{ ij} {\bf S}_j| /\sqrt{N}$, which is a quantity of 
${\cal O}(J)$. 
Therefore, similarly as we did before for $\chi_{inc}^{\prime \prime}(\omega)$,
we can perform the average over the disorder as an average over
the distribution of effective fields, to obtain the contribution
of $\chi_{high}^{\prime \prime}(\omega)$ from
\begin{equation}
\chi_{high}^{\prime \prime}(\omega) 
 \propto \int^{+\infty}_0 dh  \ h^2 P(h) 
\chi_{o}^{\prime \prime}(\omega,h),
\end{equation}
with $\chi_{o}^{\prime \prime}(\omega,h)$ of the form (\ref{chimf})
where we point out that in the present case, the directions of
the effective fields are now {\em frozen} and thus we include the $h^2$ term
of the Jacobian of $d${\bf h}.

The integral leads to  the following functional form for this
high frequency incoherent contribution,
\begin{equation}
\chi_{high}^{\prime \prime}(\omega)= B \omega^2 \exp[-\frac{(\omega-h_0)^2}
{2 \sigma^2}].
\label{10}
\end{equation}

We were not able to determine
$h_0$ and the variance $\sigma$ of the probability distribution
analytically.  These parameters and the constant $B$
can be chosen to fit the numerical data, however, 
the sizable finite size effects observed and the small
spectral intensity did not allow us for a very 
accurate determination. 
We found that the choice of $h_0 = 1$ and $\sigma=0.5$ is consistent with the behavior 
of $\chi_{high}^{\prime \prime}(\omega)$
as the system grows to the large size limit, as shown in Fig. 8.   
We note that the spectral weight of this contribution is extremely small,
less that 5 percent of the total weight of $\chi^{\prime \prime}(\omega)$,
which implies that the agreement remains only qualitative.

A final issue that we would like to discuss briefly
concerns the difference in the estimated value for the
Edwards Anderson order parameter $q$, in comparison with the
result we reported in Ref. \cite{limarprl}. This 
has a simple explanation:
the $T=0$ sum rule $\int d\omega \chi^{\prime \prime}(\omega)=1/4$
is used to numerically compute the order parameter from the expression
\begin{equation}
q= \frac{1}{4}-\int_0^\infty d\omega  \chi^{\prime \prime}_{reg}(\omega),
\label{sum}
\end{equation} 
where $\chi^{\prime \prime}_{reg}(\omega)$ denotes the regular part of
the spectral function. We have already discussed the many
different contributions to the $\chi^{\prime \prime}_{reg}(\omega)$,
including the small last one from $\chi_{high}^{\prime \prime}(\omega)$, which went unnoticed
in our previous work \cite{limarprl}. The spectral weight carried by
$\chi_{high}^{\prime \prime}(\omega)$ is only  $\sim 0.015$ and is in fact 
the origin and amount
of the difference.
We have now achieved a better and more accurate description of the regular part of the
frequency dependent spin-spin response function that through (\ref{sum}) enabled us 
to obtain a more accurate estimate for the order parameter $q$.

\subsection{$\chi^{\prime \prime}_{loc}(\omega)$ at finite $T$.}

We now turn to the behavior of the dynamic response at finite temperature.
We recall the definition of the spectral density $\rho(\omega)$,
\begin{equation}
\rho(\omega) = \frac{1}{e^{-\beta \omega}-1}
\chi^{\prime \prime}_{loc}(\omega). 
\end{equation}
It is more convenient to use
$\rho(\omega)$ instead of $\chi^{\prime \prime}_{loc}(\omega)$.
While both quantities coincide at $T=0$,
the former obeys a simple sum-rule which allows for a clearer 
interpretation of the transfer of spectral weight. In fact,
the sum-rule reads
\begin{equation}
\int_0^\infty \rho(\omega) d\omega = \frac{1}{4}.
\end{equation}

The heuristic derivation of a response function for the incoherent
degrees of freedom  that we presented before 
can be easily extended to finite temperatures.
We obtain
\begin{equation}
\chi^{\prime \prime}_{inc}(\omega,T) = 
C \exp{[-\frac{\omega^2}{2 J S(S+1)}]} tanh(\frac{\beta\omega}{2}).
\label{e20}
\end{equation}
and correspondingly,
\begin{equation}
\rho_{inc}(\omega,T) =
C \exp{[-\frac{\omega^2}{2 J S(S+1)}]} tanh (\frac{\beta\omega}{2}) 
\frac{1}{e^{-\beta \omega}-1} .
\end{equation}
where the constant $C$ is now $C=C(T)$.
Rather remarkably,  the 
analysis of our results demonstrate that the functional form for
$\rho_{inc}(\omega, T)$ remains an excellent fit 
for the incoherent part of the spectra. Moreover, we found that within the
temperature range investigated ($0$ to $\sim 2T_g$) the numerical value of the
coefficient $C$ remains a constant within our numerical precision.
We should mention, however, that at high enough temperature, at least the numerical
value of $C$ has to change for the fit to comply with the sum-rule. The rather 
uninteresting high temperature regime has been left out from the scope of the present work.

Similarly as we did before, we use the expressions above to guide our analysis
of the numerical data. We define 
\begin{equation}
\delta \rho(\omega, T) =
\rho(\omega, T) - \rho_{inc}(\omega, T).
\label{deltarho}
\end{equation}
Typical curves for $T=0.25$ are depicted in Fig. 9 for different 
system sizes. As in the case of $T=0$ the response is split 
in a low frequency feature and a higher frequency one. The low frequency peak 
corresponds to the $\delta$-function contribution discussed at $T=0$ that now 
acquires a {\em finite} width (${\cal O}(T)$)
due to the slow diffusive modes that appear at finite 
temperatures.

A last and smaller contribution is also apparent at higher frequencies of
${\cal O}(J)$. It is nothing but the signature of the high frequency
contribution that we discussed at the end of the previous subsection. It
corresponds to the fast precession of single unbinded spins around the now slowly
varying magnetic fields. While the physical picture is qualitatively clear, we
were  not able to perform reliable extrapolations to the thermodynamic limit to
extract quantitative detailed results upon the evolution of the spectrum as $T$ increases.
In particular, it was not possible to obtain an accurate  
estimate of the Edwards-Anderson order parameter as a function of $T$. 
While no technical impediment exists a priori with the method, it turns out that 
the statistics that we collected were not sufficient. These details will have
to wait for longer runs or more powerful computational resources. 
A final point we would like to consider is the transfer of
spectral weight. Although the size of our systems is limited
to 12 sites, this seems to be sufficient for a qualitative
discussion the evolution and transfers of spectral
weight of $\rho(\omega)$ as a function of temperature.

In Fig. 10 we show results for $\delta \rho(\omega, T)$ 
at temperatures in the range of 0.04 to 0.25. One may think of this
contribution to the spectra as due to the slow degrees of freedom
that were originally frozen at $T=0$ and then gradually melt as $T$ is increased.
Our results indicate that the width of the low frequency peak
is linear in $T$, as may be expected for diffusive modes \cite{foster}. 
Interestingly, the small higher frequency part is a rather broad feature
that does not show any significant shift nor broadening with temperature,
but merely looses spectral intensity with increasing $T$. At 
finite $T$, the effect of the gradual melting of the clusters of frozen spins  
leads to a slow (difussive) motion of the directions in which they
point. However,  as long as the drift of those clusters is slow compared to the
fast precession of the unbinded spins, the higher frequency contribution should remain
approximately unchanged, except for its spectral intensity which decreases
with $T$. 

This brings us to a last issue worth mentioning, namely,
the transfer of
spectral weight towards the background when $T > T_g$.
This feature is better seen in the integrals
\begin{equation}
I(\omega,T)=\int_{-\infty}^{\omega} d \omega^{\prime} 
\rho(\omega^{\prime},T),
\label{int}
\end{equation}
which are shown in Fig. 11 for the same set of temperatures as in Fig. 10.

It is clear in Fig. 11 that as the temperature increases,
the distinction between low and high degrees of freedom
gradually disappears and that the width of the main feature grows towards
the single bare energy scale of the model $J$ that must control
the physics at high temperatures.

\section{The specific heat.}
In real materials with a spin-glass phase below $T_g$,
the specific heat $C_v$ exhibits a smooth maximum well above the critical
temperature \cite{fisher,young,binder}. The origin of this behavior
remains an open question as it cannot be explained from the predictions of  
classical models. In particular, for the Sherrington
Kirpatrick model \cite{sk}, which is the classical version of the Heisenberg
model, it is found that $C_v$ has a maximum exactly at $T^{SK}_g$
with a small discontinuity in $\partial C_v / \partial T$
 \cite{binder}.  
In this section we shall argue that the observed behavior of $C_v$ can be
understood as a result of quantum effects. We shall discuss the role of 
quantum fluctuations in the behavior of the specific heat 
\begin{equation}
C_v=\frac{\partial E}{\partial T}.
\label{11}
\end{equation}
and show that it leads to a shift of the maximum of $C_v$ towards temperatures
higher than $T_g$.

The mean value of the energy per spin $E$ is computed from
\begin{eqnarray}
E&=&\frac{1}{N}  
\frac{1}{M} \sum_{m=1}^{M} \frac{1}{Z^{(m)}} \sum_k [
e^{-\beta E_k^{(m)}}
 E_k^{(m)}] \nonumber \\
&=&\int_{-\infty}^{+\infty} d\omega \  \omega \chi^{\prime \prime}_{loc}(\omega),
\label{12}
\end{eqnarray}  
and (\ref{11}) is obtained by numerically differentiating this 
quantity.
The energy $E$ as a function of $T$ is shown in Fig. 12 for several system
sizes. 
In the spirit of the previous analysis of the different contributions
to $\chi^{\prime \prime}_{loc}(\omega)$ and using
expression (\ref{12}), we may think
of $E$ as resulting from two different contributions: $E=E_{inc}+E_{coh}$.
The energy $E_{inc}$ comes
from the fraction of degrees of freedom that remain unfrozen down to $T=0$
due to quantum fluctuations, and $E_{coh}$ is the gain in energy due to
the formation of the glass.
As it turns out, the former is the largest contribution of the two and is obtain
by replacing (\ref{e20}) in (\ref{12})
\begin{equation}
E_{inc}= C \int_{-\infty}^{+\infty} dh  \ h
\exp[\frac{-h^2}{2 J^2 S(S+1)}] tanh (\frac{\beta h }{2}).
\label{13}
\end{equation}
This energy is indicated in thick line in Fig. 12.
At low temperatures it underestimates (in absolute value)
the total $E$ as it lacks the sizeable contribution from the
frozen degrees of freedom. On the other hand, at higher temperatures its qualitative
behavior is similar to that of $E(T)$.

The results for $C_v(T) = {\partial E}/{\partial T}$ are shown in Fig. 13
for systems of various sizes.
The contribution of the incoherent background is also plotted in thick
line for comparison. It is remarkable that  $C_v$ exhibits an smooth
cusp with a maximum at a temperature $T_M \approx 0.25$ which is
significantly larger than $T_g$. 
Interestingly, the contribution of the incoherent background 
exhibits a similar maximum at approximately $T_M$. The actual contribution
to $C_v$ coming from the frozen part is relatively small compared 
to the quantum disordered part. Thus, the latter is found to dominate
the behavior of $C_v$ and in fact it predominantly determines the actual 
position of the maximum.

Therefore, our results strongly suggest
that in the present model incoherent mean-field like excitations
due to quantum fluctuations set
the behavior of $C_v$ as a function of $T$. In real systems
the quantum fluctuations (controlled by, for instance, anisotropy)
may play a more or less important
role than in the current model but should always be present.
One may thus expect that its effect could produce 
larger or smaller shifts of $T_M$ always to higher temperatures respect
to $T_g$, as is in fact observed experimentally.

To give further support to our claims, we investigate the 
finite size effects on $C_v$ in more detail. We applied the
same numerical procedure to  
the classical Sherrington Kirpatrick model \cite{sk}, 
which is defined by the hamiltonian
\begin{equation}
H_{SK} = \frac{1}{\sqrt{N}} 
\sum_{i,j=1}^{N} J_{ij} S^z_i S^z_j.
\label{14}
\end{equation}
where $S^z_i$ denote Ising spins now.
This model has been extensively studied and many well established
results have been reported in the paramagnetic
and glassy phases \cite{fisher,young,binder}. 
Results for the behavior of $C_v$ as a function of $T$ obtained with
exact diagonalization are shown in Fig. 14 for systems of various sizes.
Although these data does not allow us to capture a subtle issue such as
the small cusp in  $ C_v$  at $T^{SK}_g$ in the thermodynamic limit,
it is clear that the curves become sharper as the size of the
system increases. More remarkable is the fact that for these finite
systems the maximum $T_M$ occurs {\em below} $T_g$,
in marked contrast with the behavior of $C_v$ in the random Heisenberg model
that is also depicted in the figure for comparison. 
Note that the finite size effects on $T_M$ in the quantum model are negligible.

We therefore conclude that our results show strong evidence that
the smooth maximum of $C_v$ at $T_M > T_g$  
is a genuine feature of the quantum model rather than a finite size effect.

\section{Summary and discussion}
In summary, we have studied the random infinite-range Heisenberg model
at $T=0$ and at finite temperature
with exact diagonalization techniques by carrying out a careful analysis
of the finite size effects. 
We have first shown that our approach
is able to reproduce the sole well established quantitative result known to us
in this model, namely,
the quantum Monte Carlo estimate of the
critical temperature $T_g$ \cite{daniel} for spin glass ordering. 

At $T=0$ we found that the dynamical spin response in the thermodynamic limit
behaves as $\chi_{loc}^{\prime \prime}(\omega)=q\delta(\omega)+
\chi^{\prime\prime}_{reg}(\omega)$, where $q$ is the Edwards Anderson
parameter. 
The low energy excitations producing the $\propto \delta(\omega)$
response are due to two different contributions. On one hand,
the freezing of slow collective excitations that we identified as
groups of spins that build unfrustrated clusters and coexist in
a given typical realization of disorder. The nature of these excitations is the
slow quantum tunneling between (cluster) states related by a time-reversal operation.
On the other hand, 
there is also a  second contribution to the $\propto \delta(\omega)$ response
due to a 0-mode like excitation originated in the spin-rotation
invariance of the model and the fact that there is a finite mean total spin.
Interestingly,  while the local magnetic moment of the system, i.e. the mean magnetic
moment {\em per site}, does 
vanish in the thermodynamic limit, we found that the 
total spin is $S\neq 0$ and of ${\cal O}(\sqrt{N})$.  

The regular part
$\chi^{\prime\prime}_{reg}(\omega)$ collects the contribution of incoherent
excitations due to quantum fluctuations plus the effect of high energy excitations 
due to the flipping of
some few unbinded spins. Both kinds of excitations seem to be very well
described in terms of heuristic mean-field theories that provide a qualitative
insight. 

When the temperature $T$ increases the response of the low energy excitations
is found to acquire a finite width of order $T$ due to the melting of frozen spins.
As these degrees of freedom gradually unbind and become part of the regular contribution,
the weight of the high energy excitations gets also gradually reduced.
However, the shape of this part of the response is preserved within a wide
temperature range as long as there is a separation of energy scales
between the slow and fast excitations.

We have also studied the behavior of the specific heat as a function of $T$.
We found compelling evidence that in the thermodynamic limit $C_v$ has a
smooth maximum at a temperature $T_M$, significantly larger than the transition
temperature $T_g$. This behavior is in excellent qualitative agreement with that
experimentally observed in real spin glasses and is in marked contrast with
the predictions based on classical models such as the
Sherrington-Kirpatrick hamiltonian. We have shown that the origin of
the unusual behavior can be traced to the excitations 
due to quantum disorder which dominate the behavior
of the specific heat.  

In conclusion, based in our numerical analysis, the picture we propose for the
spin-glass state is that a fraction of the system forms coexisting
unfrustrated clusters which are frozen or undergoing a slow
collective motion. The 
rest of the system remains in an uncorrelated incoherent state mostly
described by local spin dynamics.
Such a picture manifests itself in the 
different contributions to the dynamical response and particularly in
$C_v(T)$, since the incoherent excitations produce a smooth variation in the
specific heat and this quantity does not show any signature of the
paramagnetic to spin-glass transition. We find that this picture can be quite
general and that it provides a possible explanation for the  behavior of
the specific heat in real materials which remained, so far, a
standing question in the physics of spin glass systems.

Another interesting observation is that the solutions of the SU(M) generalization
of the model treated in the large M limit produces several qualitatively different
phases depending on the value of the "size" of the 
spin, $S$ \cite{subirsum,georges,parcollet}. In particular, for large $S$
the system is in a spin-glass state with a dynamic response that has a $\delta$-function
at $\omega=0$, while for $S \to 0$ the system goes into a spin-liquid state with a
response function that remains regular and goes to a constant at $\omega \to 0$.
Therefore, under the light of our results, the large M limit of the model 
seems to capture, in its different regimes, several aspects present in the M=2 solution.

From the technical point of view, we can say that the present approach offers 
a very appealing alternative for the study of disordered systems. Without the
need of approximations, it enables the treatment of paramagnetic and
disordered phases on equal footing and provides a deep insight on the nature of
the relevant excitations.
At least for the present model, we were able
to find reliable extrapolations to the thermodynamic limit for most of the
studied quantities, thus overcoming the obstacle of the finite-size effects of
the systems under study. 

Finally, many interesting issues are left for future work, such as a systematic
study of the role of dimensionality or lattice connectivity and that of 
anisotropy.

\section{Acknowledgements}
L. A. thanks G. Santoro for his comments on this manuscript.  
We acknowledge support from CONICET and from grants of ANCyPT,
Fundaci\'on Antorchas and ECOS-SECyT.

\newpage

\begin{figure}
\epsfxsize=3.2in
\epsffile{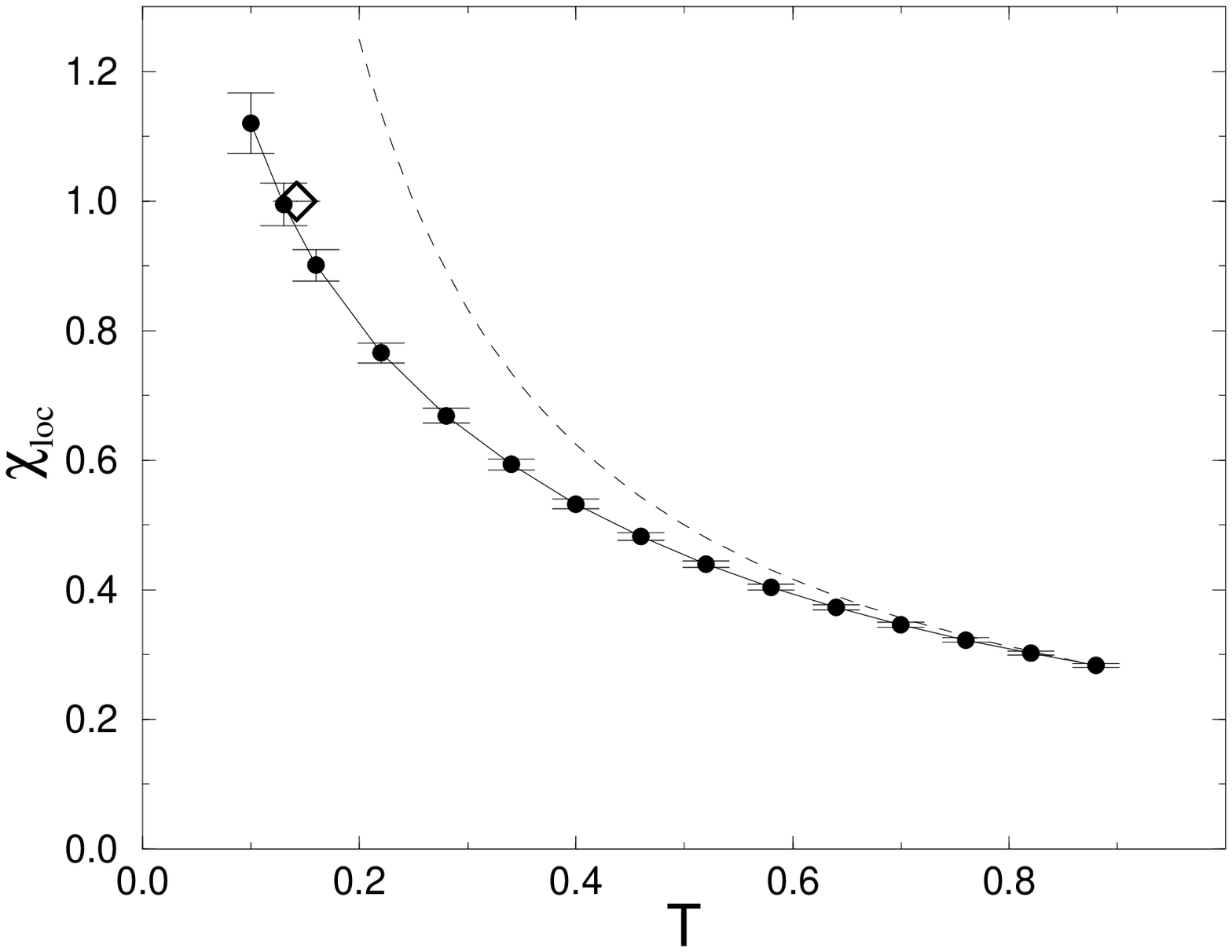}
\caption{The $\omega=0$ dynamical susceptibility $\chi_{loc}$ as a function of
$T$. Solid circles correspond to extrapolations of the numerical data to the
thermodynamic limit (error bars are indicated). 
The critical temperature $T_g$ corresponds to $\chi_{loc}=1$.
The diamond corresponds to
the result of the Monte Carlo solution of Bray and Moore equations. The
Curie law followed by classical spins is shown in dashed lines.}
\label{fig1}
\end{figure}

\begin{figure}
\epsfxsize=3.2in
\epsffile{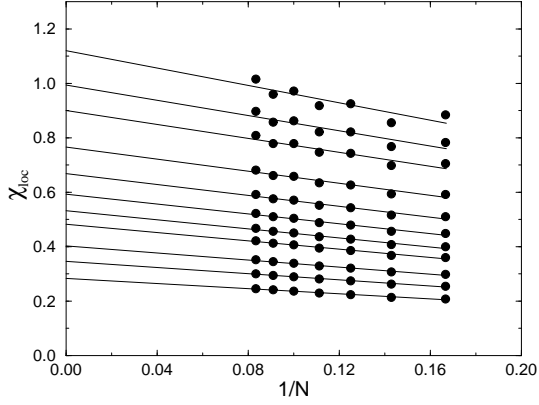}
\caption{
$\chi_{loc}$ as a function of
$1/N$ for different temperatures above and below $T_g$.
The corresponding temperatures are $T=0.1,0.13,0.16,0.22,0.28,0.34,
0.4,0.46,0.58,0.7,0.88$ (top to bottom).
 Solid
lines show the linear fitting to perform the extrapolation to the
thermodynamic limit.}  
\label{fig2}
\end{figure}

\begin{figure}
\epsfxsize=3.2in
\epsffile{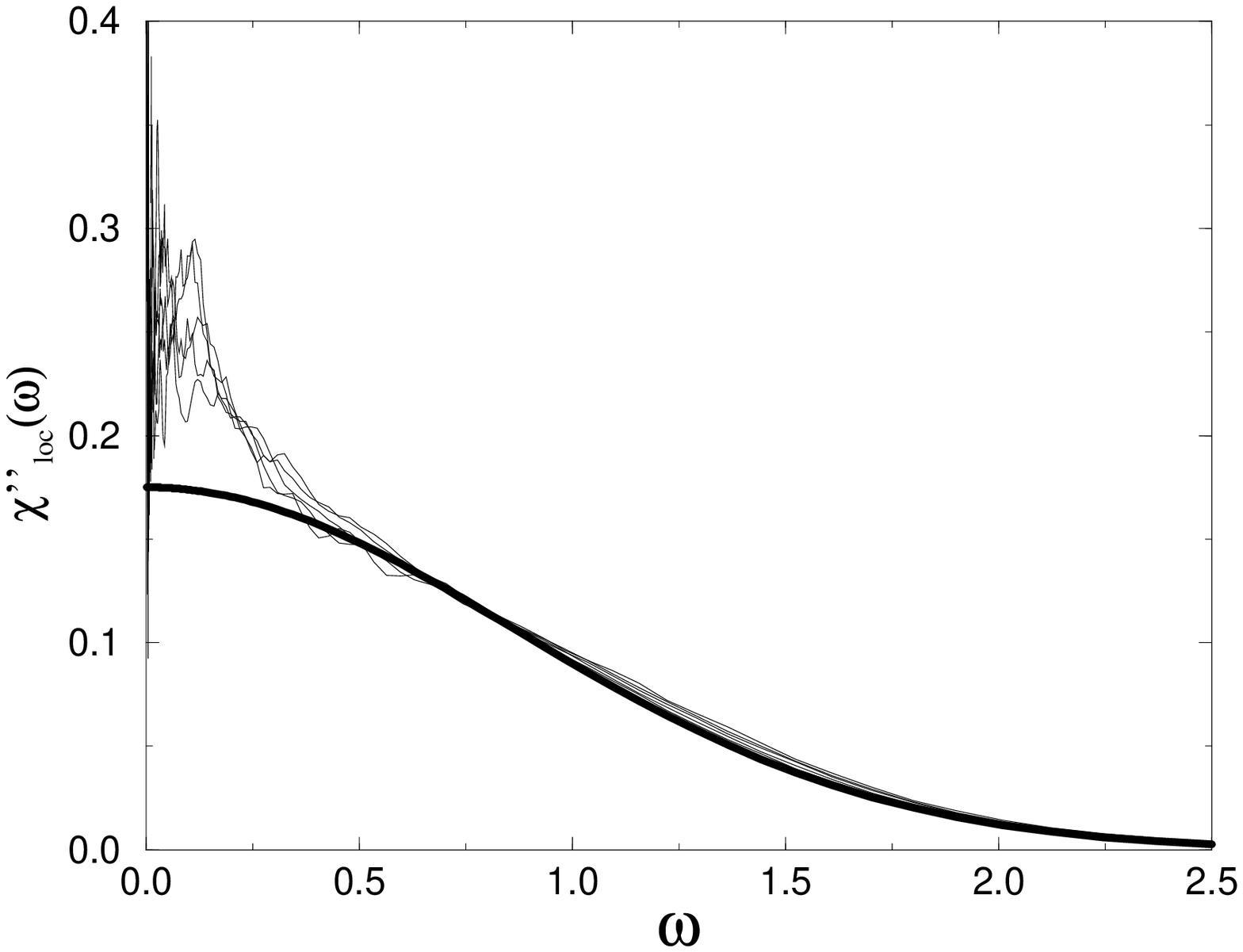}
\caption{
The spectral function $\chi^{\prime \prime}_{loc}(\omega)$ 
for systems of 8, 10, 12, 14 and 16 spins (bottom to top). 
The thick line indicates the response due to incoherent
excitations given in Eq. (\ref{7}), with $S=1/2$. The constant
$C=0.175$ is set to fit the numerical data.}
\label{fig3}
\end{figure}

\begin{figure}
\epsfxsize=3.2in
\epsffile{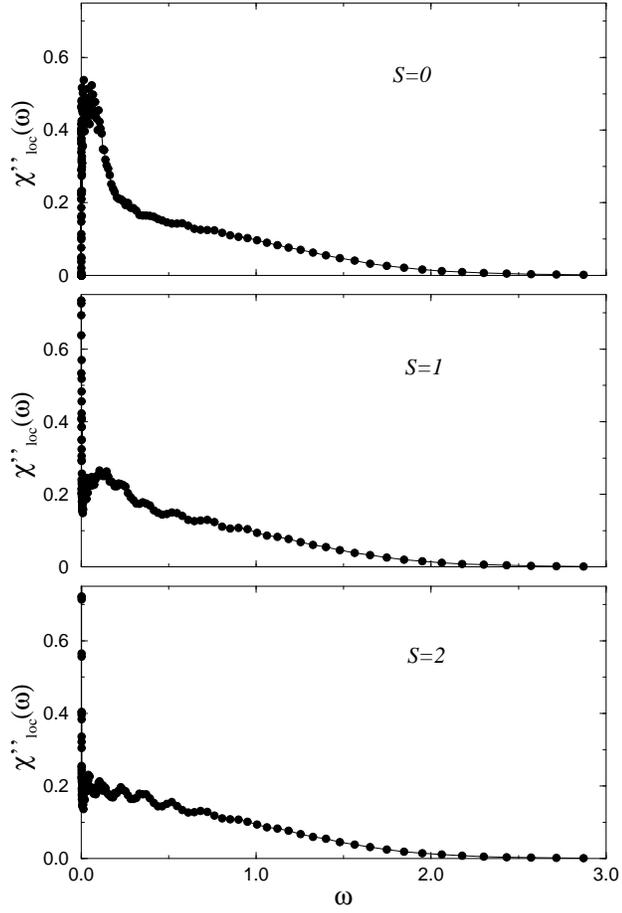}
\caption{
The spectral function $\chi^{\prime \prime}_{loc}(\omega)$ 
for $N=14$ corresponding to the
average over the set of disorder realizations with the ground state in the
$S=0,1,2$ total spin sectors (top to bottom).}
\label{fig4}
\end{figure}

\begin{figure}
\epsfxsize=3.2in
\epsfxsize=5.in
\epsffile{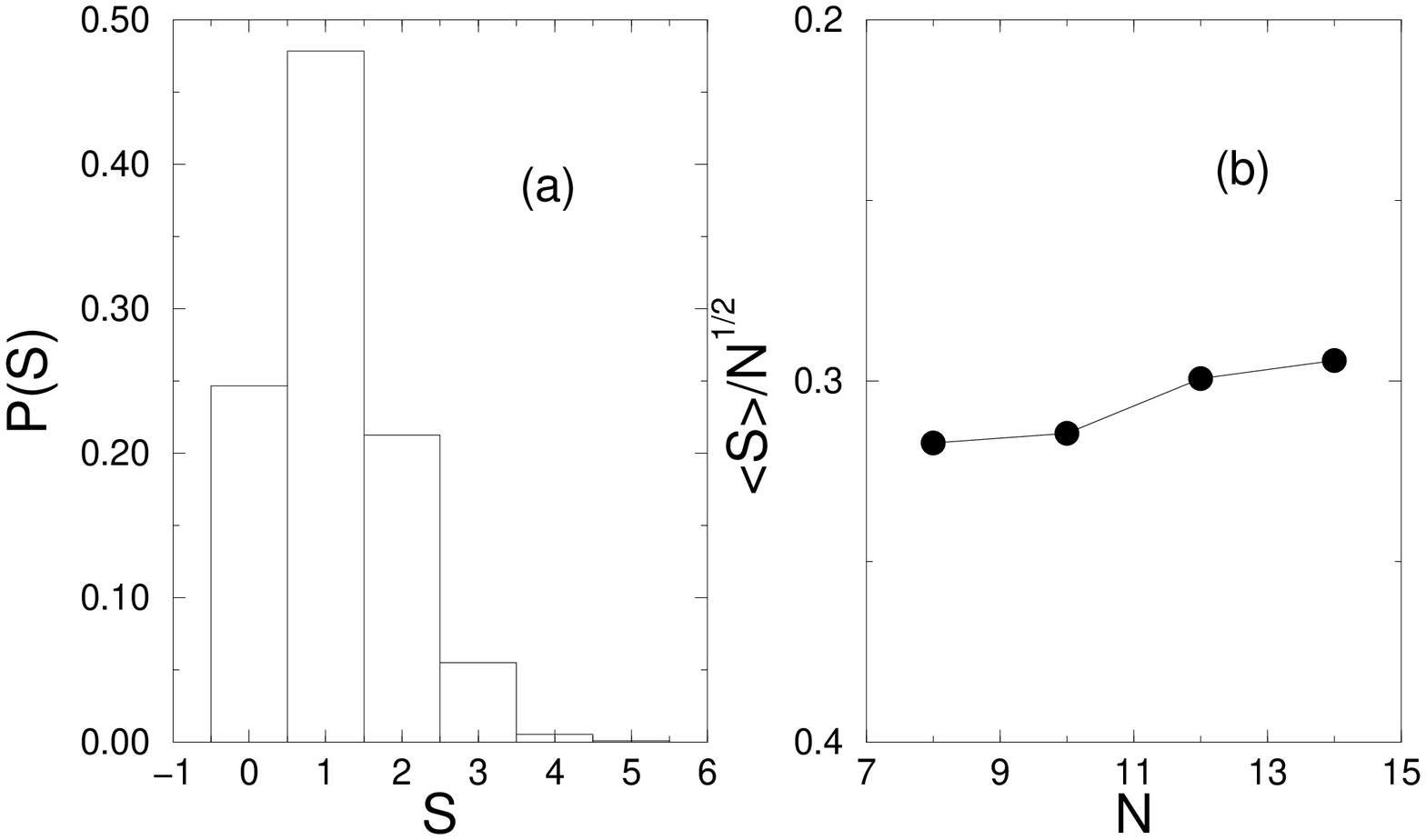}
\caption{ 
(a) Histogram for the total spin $S$ of the ground state for $N=14$. 
(b) $\langle S \rangle /\sqrt{N}$ as a function of the system size. }
\label{fig5}
\end{figure}

\begin{figure}
\epsfxsize=3.2in
\epsffile{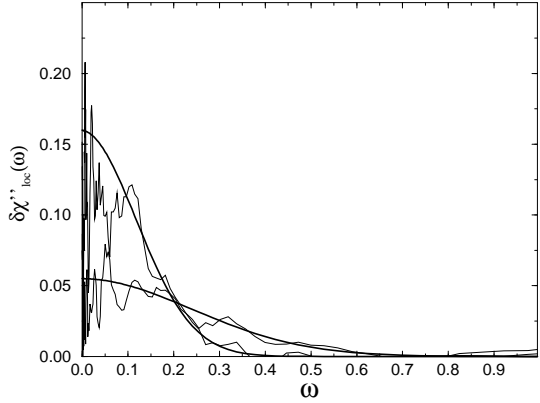}
\caption{ Details of the low frequency part of 
$\delta \chi^{\prime \prime}_{loc}(\omega)$  for systems with $N=6, 14$. 
The thick lines indicate the Gaussian fits to the numerical data.}
\label{fig6}
\end{figure}

\begin{figure}
\epsfxsize=3.2in
\epsffile{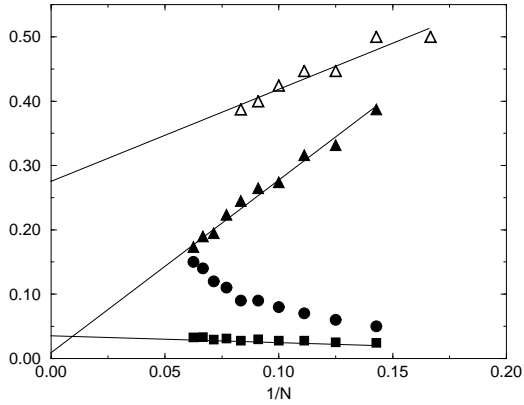}
\caption{ 
The parameters of the Gaussian fitting function for the 
low frequency contribution to the response as a function of
the inverse system size $1/N$. The circles correspond to
the height $A$ and the triangles to the width $\Gamma$ of the
Gaussian defined in Eq. (\ref{9}). The squares indicate the
spectral weight of $\chi^{\prime \prime}_{low}$.}  
\label{fig7}
\end{figure}

\begin{figure}
\epsfxsize=3.2in
\epsffile{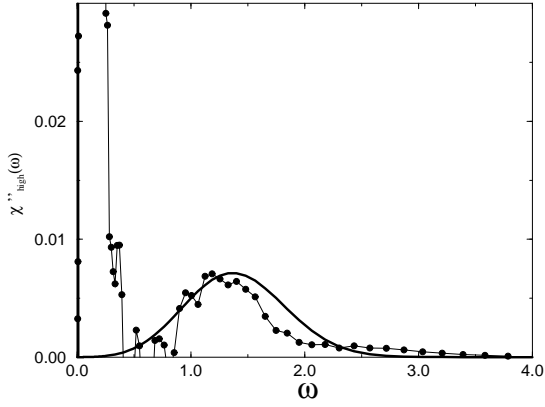}
\caption{ 
Details of the high frequency part 
of  $\delta \chi^{\prime \prime}_{loc}(\omega)$  
for a system of 14 spins. The thick line
corresponds to $\chi^{\prime \prime}_{high}(\omega)$ given by Eq.(\ref{10})
with $B=0.005$, $\sigma=1/2$ and $h_0=1$. The spectral weight of
this contribution is small, less than $5\%$ of the total spectral weight.} 
\label{fig8} 
\end{figure}

\begin{figure}
\epsfxsize=3.2in
\epsffile{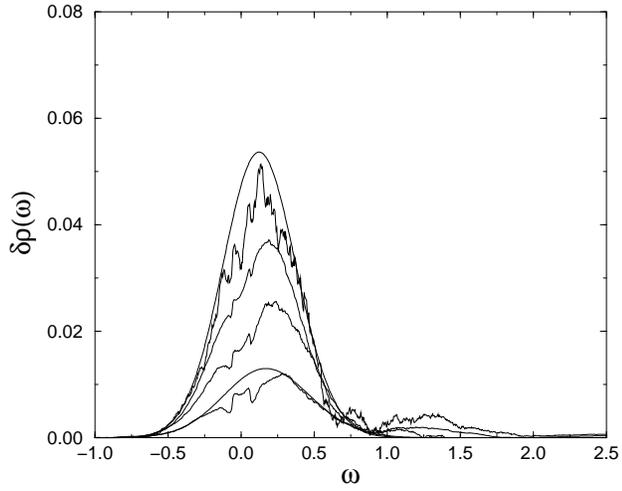}
\caption{ 
$\delta \rho(\omega,T)$  for  systems with $N=6,8,10,12$ sites at $T=0.25$.
 Thick lines are fits with Gaussians times the function $1/(1-\exp(-\beta
\omega))$.} 
\label{fig9}
\end{figure}

\begin{figure}
\epsfxsize=3.2in
\epsffile{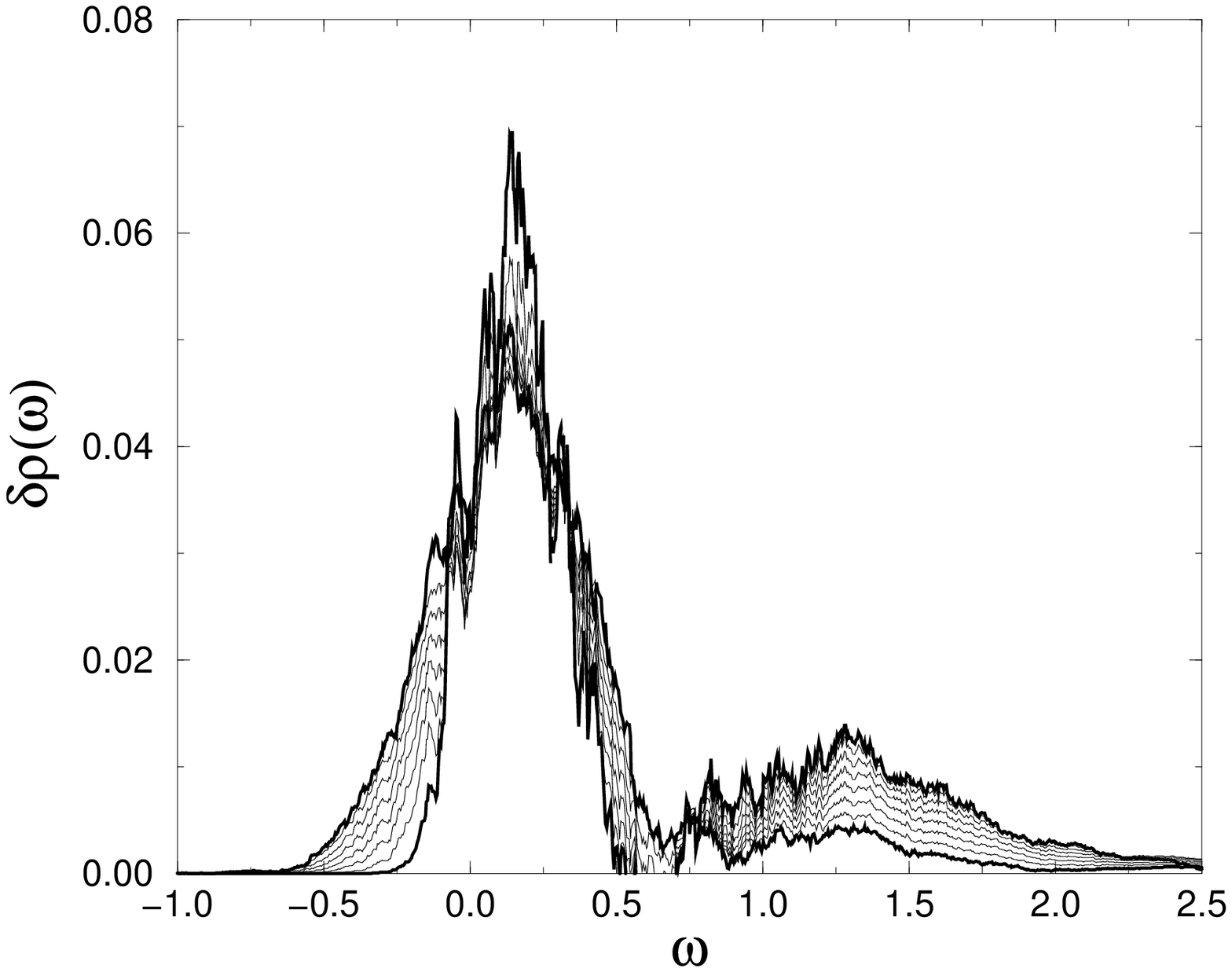}
\caption{ 
$\delta \rho(\omega,T)$  for $N=12$ and $T=0.04, 0.7, 0.1, 0.13, 0.16, 0.19,
0.22, 0.25$ (bottom to top of lower frequency feature and top to bottom 
of the higher frequency one). 
The curves corresponding to the lowest and highest temperatures
are indicated in thick lines.}
\label{fig10} 
\end{figure}

\begin{figure}
\epsfxsize=3.2in
\epsffile{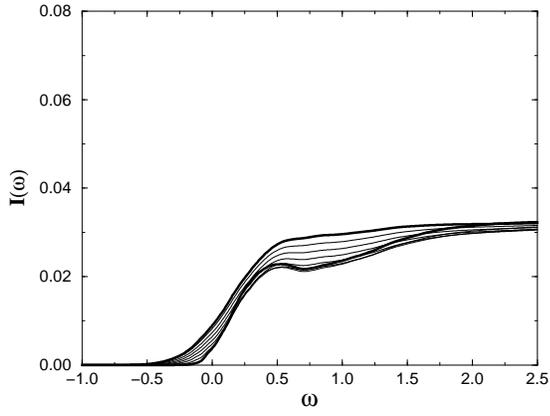}
\caption{ The integrated spectral weight $I(\omega)$ defined in (\ref{int}) 
 for $N=12$ and $T=0.04, 0.7, 0.1, 0.13, 0.16, 0.19,
0.22, 0.25$ (bottom to top). The curves corresponding to the lowest and highest temperatures
are indicated in thick lines.  
}
\label{fig11} 
\end{figure}

\begin{figure}
\epsfxsize=3.2in
\epsffile{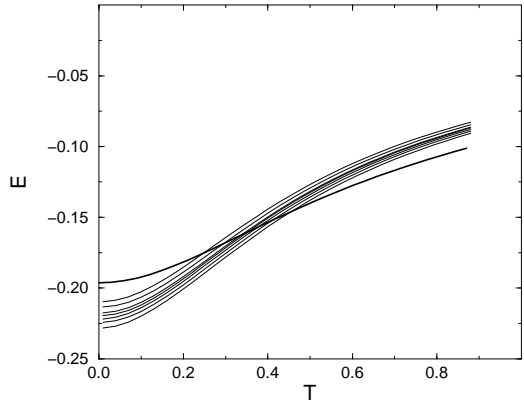}
\caption{ The mean energy $E$ defined in (\ref{12}) as  a function of
$T$ for systems with $N=6,7,8,9,10,11,12$ (top to bottom) sites.
The thick line corresponds to the contribution of the incoherent part
(\ref{13}).} 
\label{fig12} 
\end{figure}

\begin{figure}
\epsfxsize=3.2in
\epsffile{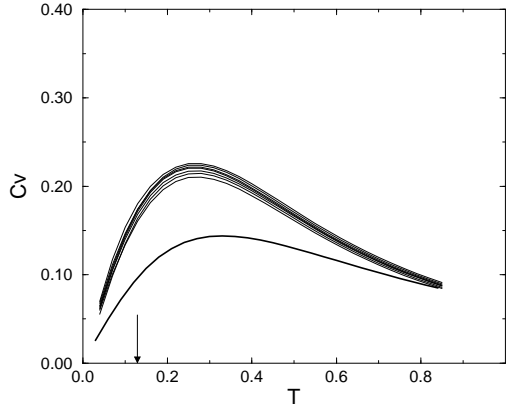}
\caption{ The specific heat $C_v$ defined in (\ref{11}) as  a function of
$T$ for systems with $N=6,7,8,9,10,11,12$ (bottom to top) sites.
The thick line corresponds to the contribution of the incoherent part.
The arrow indicates the critical temperature $T_g$.} 
\label{fig13} 
\end{figure}

\begin{figure}
\epsfxsize=3.2in
\epsffile{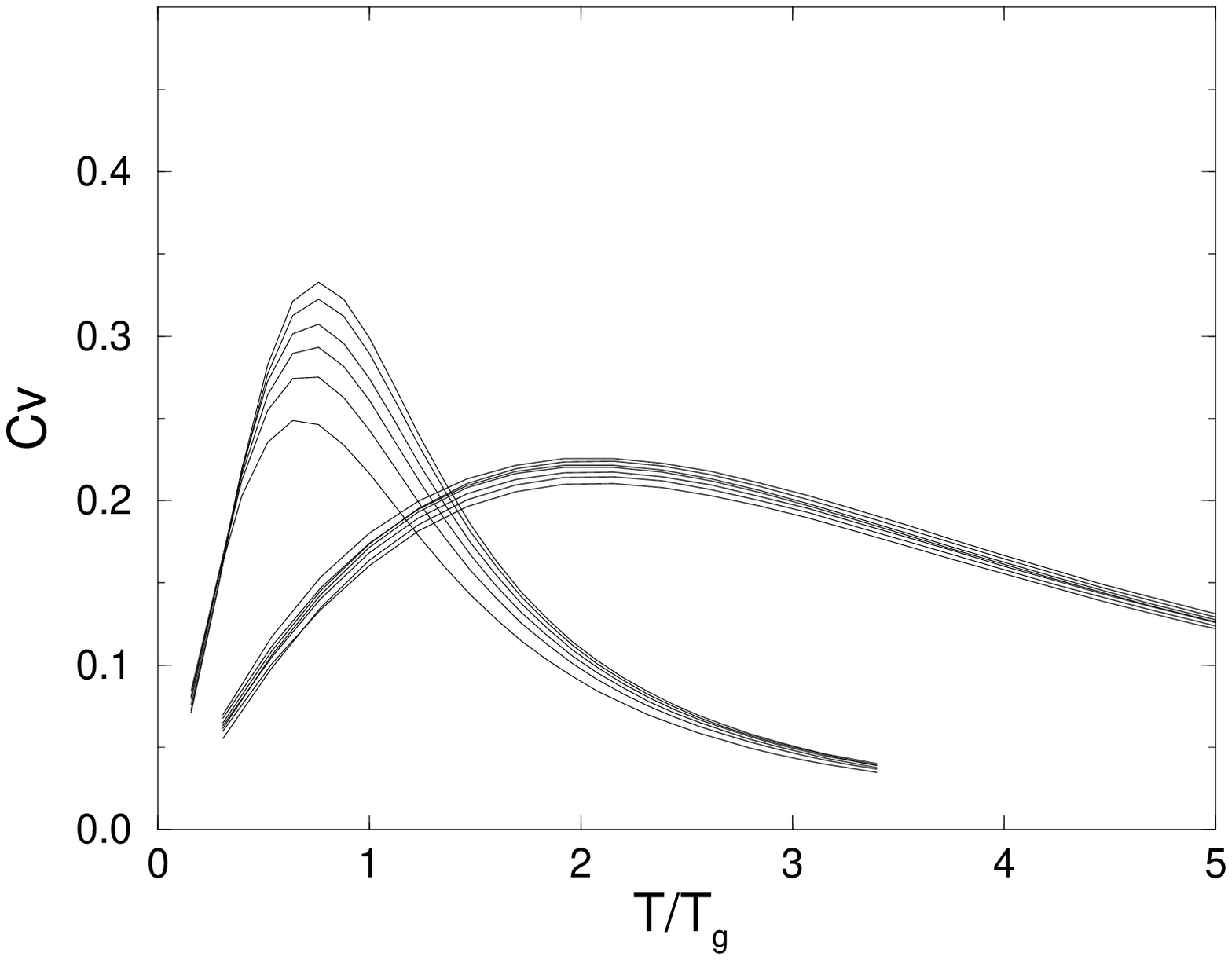}
\caption{ The specific heat $C_v$ as  a
function of $T/T_g$ for systems with $N=6,7,8,9,10,11,12$ (bottom to top)
sites. The set of curves on the right corresponds to the Heisenberg
model while the one on the left to the Sherrington Kirpatrick model (\ref{14}).
}  \label{fig14} 
\end{figure}

\newpage


\begin{references}
\bibitem{acla}  We focus our discussion to ground states with $S=0$,
adopting a basis within the $S^{z}=0$ representation. Nevertheless, our
arguments can be generalized to the other spin components.
\end{references}
\end{document}